%% file: main.tex
\documentclass[conference]{IEEEtran}
\IEEEoverridecommandlockouts

\usepackage{color}
\usepackage{cite}
\usepackage[subrefformat=parens]{subcaption}   
\usepackage{tabularx}     
\usepackage{textcomp}     
\usepackage{threeparttable} 
\usepackage{nth}          
\usepackage{gensymb}      
\usepackage{graphicx}     
\usepackage{graphics}
\usepackage{mathtools}    
\usepackage{amsmath}
\usepackage{amsfonts}
\usepackage{amsthm}
\usepackage{verbatim}
\usepackage{amssymb}
\usepackage{threeparttable}

\usepackage{stfloats}
\usepackage{setspace}
\usepackage{caption}
\usepackage{titlesec}
\usepackage{todonotes}
\usepackage{bbm}

\newcommand{\y}{\mathbf{y}}

\newcommand{\x}{\mathbf{x}}
\newcommand{\s}{\mathbf{s}}

\usepackage[linesnumbered,vlined,ruled]{algorithm2e}

\begin{document}

\title{Deep Learning Based Near-Orthogonal Superposition Code for Short Message Transmission}

\author{ Chenghong Bian, Mingyu Yang, Chin-Wei Hsu, Hun-Seok Kim,~\IEEEmembership{Member,~IEEE}  \\
Department of EECS, University of Michigan, Ann Arbor, MI \\
\{chbian, mingyuy, chinweih, hunseok\}@umich.edu}

\maketitle

\begin{abstract}
Massive machine type communication (mMTC) has attracted new coding schemes optimized for reliable short message transmission. In this paper, a novel deep learning based near-orthogonal superposition (NOS) coding scheme is proposed for reliable transmission of short messages in the additive white Gaussian noise (AWGN) channel for mMTC applications. Similar to recent hyper-dimensional modulation (HDM), the NOS encoder spreads the information bits to multiple near-orthogonal high dimensional vectors to be combined (superimposed) into a single vector for transmission. The NOS decoder first estimates the information vectors and then performs a cyclic redundancy check (CRC)-assisted $K$-best tree-search algorithm to further reduce the packet error rate. The proposed NOS encoder and decoder are deep neural networks (DNNs) jointly trained as an auto-encoder and decoder pair to learn a new NOS coding scheme with near-orthogonal codewords. Simulation results show the proposed deep learning-based NOS scheme outperforms HDM and Polar code with CRC-aided list decoding for short (32-bit) message transmission. 
\end{abstract}


\IEEEpeerreviewmaketitle

\input{introduction.tex}

\input{dl_model.tex}

\input{kbest.tex}

\input{evaluation.tex}
\vspace{-1.0mm}
\section{Conclusion}
\vspace{-1.0mm}

This paper proposes a novel deep learning based NOS coding scheme for short packets. The proposed framework enables the encoder to successfully learn near-orthogonal codewords. The receiver uses the near-orthogonality property to realize a MAP algorithm while the error performance is further enhanced by $K$-best decoding with CRC assistance. Simulation results show the proposed scheme outperforms HDM and Polar code with lower PER for short message transmission in the AWGN channel.

\vspace{-1.0mm}
\section*{Acknowledgement}
\vspace{-1.0mm}
This work was funded in part by DARPA YFA \#D18AP00076 and NSF CAREER \#1942806.

\bibliographystyle{IEEEtran}
\bibliography{IEEEabrv,ref}

\end{document}

%% file: introduction.tex
\section{Introduction}


Massive machine type communication (mMTC) is expected to play an essential role for next generation wireless standards with a wide range of applications including health, security and transportation \cite{mMTC2}. These applications, by nature, typically employ short messages/packets carrying a relatively small number of information bits, which makes conventional codes designed for large block length less effective with high error floor or non-negligible coding gain loss.
Polar codes with list decoding \cite{polarld} is proven to be more reliable compared with other modern codes such as LDPC and Turbo code under short block lengths \cite{compare_codes}. However, its performance is far from capacity and thus new coding schemes has been actively investigated for short packets \cite{Coskun2019}. 

Hyper-dimensional modulation (HDM) is a recently proposed non-orthogonal modulation scheme for short packet communications \cite{HDM2018, HDM2019, HDMkbest}. HDM can be seen as a joint coding-modulation method and a type of superposition codes \cite{SPARC2014}.
Instead of combining selected codewords from a random codebook as in typical superposition codes, HDM uses fast Fourier transformation (FFT) and pseudo-random permutations to encode sparse pulse position modulated information vectors to a non-sparse superimposed hyper-dimensional vector for efficient encoding and decoding.
HDM was first proposed with a demodulation algorithm using an iterative parallel successive interference cancellation (SIC) technique \cite{HDM2018}. It is then extended using a $K$-best decoding algorithm \cite{HDMkbest} in AWGN and interference-limited channels to outperform the state-of-the-art CRC-assisted Polar code \cite{polarld} applied to binary phase-shift keying (BPSK) under the same spectral efficiency. Despite its excellent reliability and low complexity for short message packets, 
FFT and permutation based codeword generation in HDM is sub-optimal as a superposition code.

Recently, deep learning with deep neural networks (DNNs) has been applied to the realm of channel coding \cite{deepdec,deepae,deepldpc,LEARN,jiang2019turbo} and  further extended to joint source channel coding (JSCC) \cite{JSCC,jsccofdm}.  DNNs are first applied to decode linear block codes to replace hand-crafted (computationally-demanding) decoding algorithms for Polar and LDPC while the encoder is unchanged. Taking advantage of powerful deep learning, \cite{deepdec} and \cite{deepldpc} show improved decoding performance and enhanced robustness to various channel conditions. Meanwhile, new codes have been investigated via end-to-end deep learning. A DNN-based learned code was originally introduced in \cite{deepae} where the encoder learns a joint coding and modulation scheme generating a length-7 codeword from a length-16 one-hot input to achieve the performance similar to (7,4) hamming code. However, due to \textit{the curse of dimension}, this scheme does not scale well to longer block lengths since the dimension of the one-hot input grows exponentially with block length to make the DNN training practically infeasible. To address this issue, \cite{LEARN} proposes a RNN-based auto-encoder that emulates a conventional convolution code (CC) which takes bit input instead of processing one-hot encoded input vectors. This \textit{learned} CC outperforms conventional CC in terms of BER. In \cite{jiang2019turbo}, the authors propose a learned turbo auto-encoder where the encoder is comprised of CNNs with interleaving blocks and the decoder unfolds the iterative decoding process to multiple DNN layers to achieves the BER performance comparable to the conventional turbo code.

Inspired by HDM and prior DNN-based coding schemes, we introduce a new DNN-based near-orthogonal superposition (NOS) coding scheme to learn near-orthogonal codewords for superimposed transmission for short packets. Our learned NOS scheme takes input information bits in the form of multiple independent one-hot vectors. These information vectors are jointly encoded by a DNN and added (i.e., superimposed) together for transmission. The transmit message includes cyclic redundancy check (CRC) bits as a part of information bits to enhance the reliability in low signal-to-noise ratio (SNR) scenarios. A DNN based decoder is trained to estimate the probability of superimposed one-hot coded information vectors from the received vector. As a post-processing step, we apply a $K$-best decoding algorithm to generate $K$ candidate bit sequences with CRC validation to identify the correct one. 

The main contributions of this paper can be summarized as 1) a new deep learning based near-orthogonal superposition (NOS) code specifically designed for reliable short packet transmission, 2) a new CRC-assisted $K$-best decoding algorithm to significantly improve the PER performance beyond the capability of the DNN-based decoder, and 3) numerical evaluations to quantify the gain of the learned NOS scheme compared to HDM and Polar code which are known to be very robust for short packets.

%% file: dl_model.tex
\vspace{-1mm}
\section{NOS Code Learning}
In this section, we briefly recap superposition coding and then propose a new DNN based NOS coding.

\begin{figure*}[ht]
\centering
\includegraphics[width=0.85\linewidth,height=150pt]{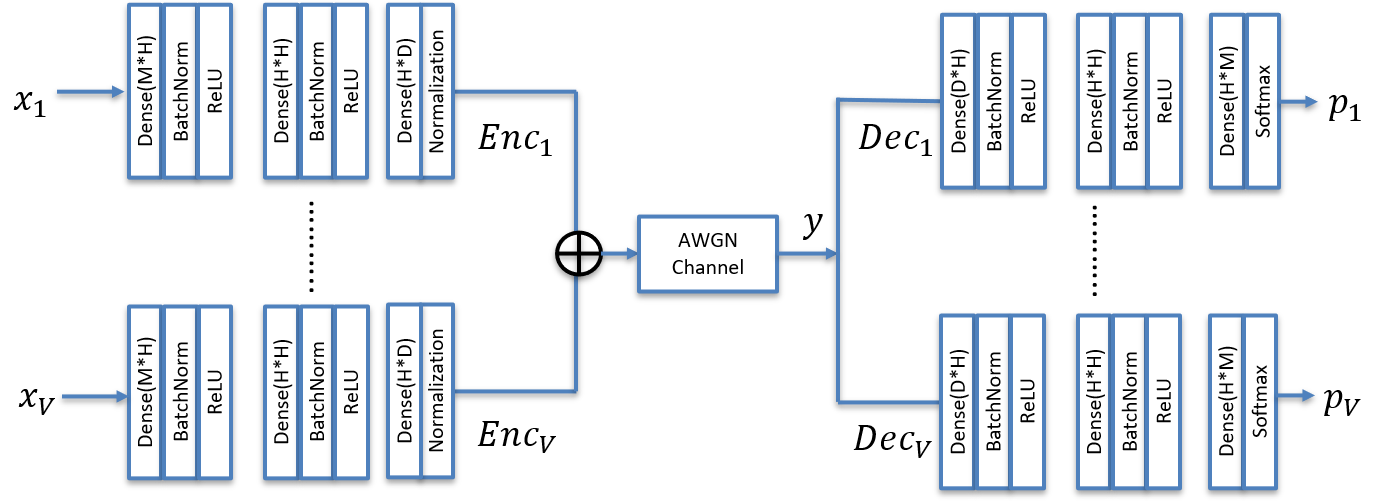}
\vspace{-0.1cm}
\caption{The neural network architecture of the proposed NOS code. }
\label{fig:network}
\end{figure*}

\vspace{-2mm}
\subsection{Superposition Coding}
Consider a sequence of independent information bits $\mathbf{b}$ whose length is $V\times m$ bits. It is split into $V$ smaller bit sequences $\mathbf{b}_i, i = 1,\cdots,V$ each carrying $m$ bits. These $\mathbf{b}_i$ are transformed to $V$ one-hot vectors $\mathbf{x}_i$ with length $M = 2^m$ whose only non-zero position (with value 1) is determined by $\mathbf{b}_i$. 
A superposition code is defined by a real-valued codebook $\cal C$ with dimension $V\times M \times D$ where 
$D$ is the codeword length. The codeword corresponding to the bit sequence $\mathbf{b}_i$ or the equivalent one-hot vector $\mathbf{x}_i$ is obtained by $\x_i {\cal C}[i,:,:]$. The superimposed transmit vector $\mathbf{s}$ with length $D$ for the entire bit sequence $\mathbf{b}$ is then obtained by adding (superimposing) $V$ codewords such that:
\vspace{-2mm}
\begin{align}
\mathbf{s} = \sum_{i=1}^V \x_i {\cal C}[i,:,:].
\end{align}

There exist efficient decoding algorithms for superposition codes including success interference canceling (SIC) as in \cite{HDM2018,SPARC2014} and approximate message passing (AMP) proposed in \cite{SPARC_AMP}. The performance of superposition code is determined by the codebook $\cal C$ and the decoding scheme. Although it is known that a random codebook $\cal C$ with CRC decoding performs reasonably well, it is difficult to find an optimal codebook jointly designed with a decoding scheme. This challenge inspired us using DNNs to jointly learn a codebook as well as a decoding scheme in an end-to-end manner as introduced in the next section.

\vspace{-1mm}

\subsection{Learned NOS Code}
The DNN architecture of the learned NOS code encoder and decoder is shown in Fig. \ref{fig:network}. The transmitter is comprised of $V$ separate encoders denoted as $Enc_i, i \in [1,V]$ and the receiver contains the corresponding $Dec_i$. Note that the network size is governed by the number of activations $H$ in the hidden layer, which is set to $H = 4D$ in our experiments.

For the encoding process, an one-hot vector $\mathbf{x}_i$ is fed to the corresponding $Enc_i$ to generate a real-valued coded vector $\mathbf{s}_i=Enc_i(\x_i)$ of length $D$. 
Since each $\mathbf{s}_i$ conveys the same amount of information and is independent, we assign the same energy $\mathbf{s}_i^*\mathbf{s}_i=\frac{D}{V}$ to each $\mathbf{s}_i$ by using a power normalization layer at the end of each $Enc_i$. The transmitted signal $\mathbf{s}$ is obtained by adding  all $\mathbf{s}_i, i=1, 2, ... ,V$:
\begin{align}
\mathbf{s} = \sum_{i=1}^V Enc_i(\x_i).
\end{align}

Instead of transmitting a real-valued signal, we convert the length-$D$ real-valued vector $\mathbf{s}$ into a complex vector $\tilde{\mathbf{s}}$ to improve spectral efficiency by taking the first $D/2$ (assume $D$ is even) elements of $\mathbf{s}$ as the real part and the rest as the imaginary part. The received signal $\tilde{\y}$ in an AWGN channel:
\begin{align}
\tilde{\y} = \tilde{\s} + \mathbf{n},
\label{eq:awgn_model}
\end{align}
where $\mathbf{n} \sim \mathcal{CN}(0,N_0\mathbf{I})$ is the complex Gaussian noise vector with element-wise variance $N_0$. At the receiver, $\mathbf{\tilde{y}}$ is first converted to a real-valued vector $\mathbf{y}$ of length $D$ and then fed into each $Dec_i$ which produces/estimates the probability vector $\mathbf{p}_i = Dec_i(\y)$. The length of $\mathbf{p}_i$ is $M$ and $\mathbf{p}_i[m]$ represents the probability of $\x_i[m]=1$.

The training of the NOS code encoder and decoder is performed using the
cross-entropy loss for each pair of the probability vector $\mathbf{p}_i$ and the one-hot input $\mathbf{x}_i$. Since $\mathbf{x}_i$ and $\mathbf{x}_j, i \neq j,$ are independent, the total loss is the summation of pairwise losses:
\begin{align}
loss = -\sum_{i=1}^V {\sum_{m=1}^{M}{\mathbf{x}_i[m] \log(\mathbf{p}_i[m])}}.
\label{eq:loss}
\end{align}
We use an ADAM optimizer to train the proposed DNN.

\vspace{1mm}
\section{Properties of Learned Codebook}
In this section, we inspect the properties of the learned NOS code and propose a simplified decoding scheme.

\subsection{Near-orthogonal Codewords}
After training, the learned codebook ${\cal C}$ with dimension $V \times M \times D$ is obtained by enumerating all length-$M$ one-hot vectors for each encoder:
\begin{align}
{\cal C}[v,m,:] = Enc_v(\x_m).
\label{eq:loss}
\end{align}


We first examine the inner product between codewords to form a codeword cross-correlation matrix $corr$ with dimension $(V,V-1,M,M)$ which is defined as:
\begin{equation}
    \begin{split}
    Corr[i,j,k,l] &= \frac{|{\cal C}[i,k,:]^T {\cal C}[j,l,:]|}{D/V} \\ 
    i,j \in [1,V]; & i \neq j ; k, l \in [1,M] 
    \end{split}
    \label{eq:inter_corr}
\end{equation}

\begin{figure}[t]
\centering
\includegraphics[width=\columnwidth]{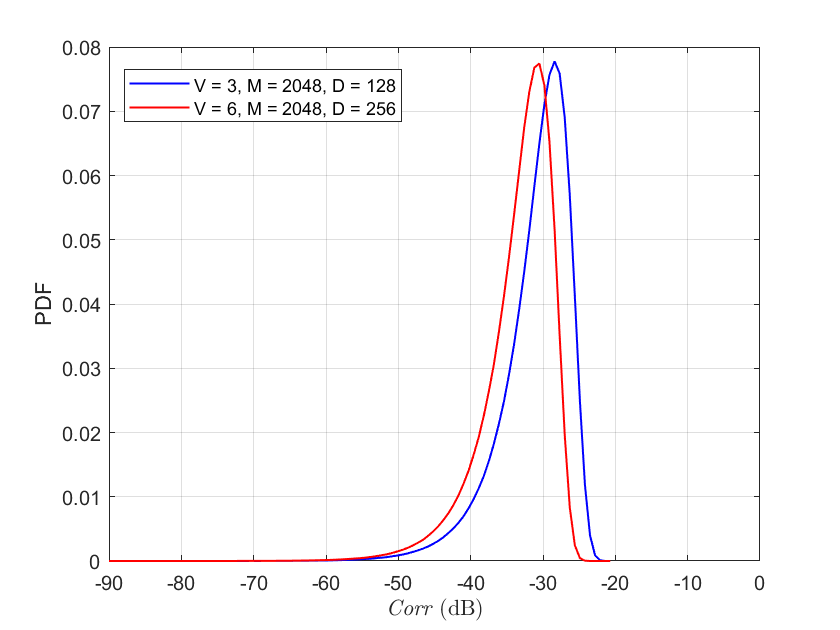}
\vspace{-0.3cm}
\caption{The distribution of $Corr$ with system parameter $(V = 3, M = 2048, D = 128)$ and $(V = 6, M = 2048, D = 256)$.}
\label{fig:inter_corr}
\end{figure}

Fig. \ref{fig:inter_corr} shows the distribution of entries in $Corr$ given $(V=3, M=2048, D=128)$  and $(V = 6, M = 2048, D = 256)$, which confirms the learned codewords are \textit{near-orthogonal} to each other. The maximum cross-correlation is approximately $30$dB lower than the energy of each codeword ($D/V$). 

\subsection{MAP Decoder}
Assuming the learned codewords are orthogonal, we derive a maximum a-posteriori probability (MAP) decoder. For MAP derivation, $\mathbf{x}_v^{m_v}$ denotes the one-hot vector $\mathbf{x}_v$ whose non-zero position is $m_v$ so that $\mathbf{s}_v^{m_v} = Enc_v(\mathbf{x}_v^{m_v})$ holds. Applying the Bayesian rule with equally probable $\mathbf{x}_v^{m_v}$, and combining the orthogonality assumption and normalized energy we obtain:
\begin{equation}
    \begin{split}
    P(\mathbf{x}_v^{m_v}&|\y) = \frac{ \sum_{i=1, i \neq v}^V \sum_{m_i=1}^M{P(\y|\mathbf{x}_1^{m_1},\cdots, \mathbf{x}_V^{m_V})}}{M^{V}P(\y)}  \\
    &\propto  exp\{\frac{2\y^T\mathbf{s}_v^{m_v}}{N_0}\} \sum_{i=1, i \neq v}^V \sum_{m_i=1}^M exp\{\frac{{2\y^T\mathbf{s}_i^{m_i}}}{N_0}\}\}.
    \label{eq:MAP}
    \end{split}
\end{equation}
Using the property that the summation over all possible encoded vectors $\mathbf{s}_i^{m_i}, i \neq v$ is constant and independent with $\mathbf{s}_v^{m_v}$, eq. \eqref{eq:MAP} can be simplified to:
\begin{align}
    P(\mathbf{x}_v^{m_v}|\y) &\propto \quad {exp\{\frac{2\y^T\mathbf{s}_v^{m_v}}{N_0}}\}.
    \label{eq:proportion_prob}
\end{align}
In this simplified MAP decoding scheme, the receiver only needs to calculate $exp\{\frac{2\y^T\mathbf{s}_v^{m_v}}{N_0}\}$ for every encoded vector in the codebook $\cal C$ to estimate a-posteriori probability:
\begin{align}
    P(\mathbf{x}_v^{m_v}|\y) = \frac{exp\{\frac{2\y^T\mathbf{s}_v^{m_v}}{N_0}\}}{\sum_{m=1}^{M}{exp\{\frac{2\y^T\mathbf{s}_v^{m}}{N_0}\}}}.
    \label{eq:final_prob}
\end{align}

\begin{figure}[t]
\centering
\includegraphics[width=\columnwidth]{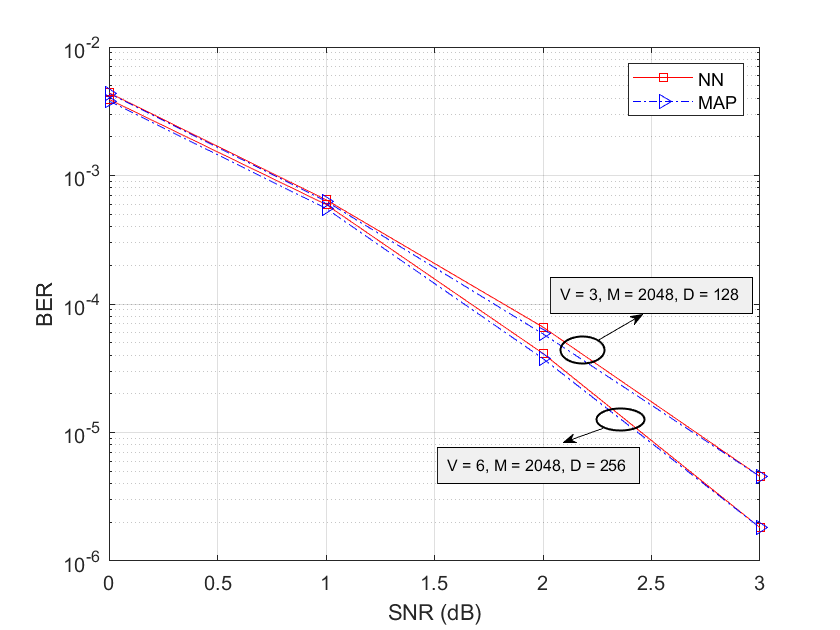}
\vspace{-0.3cm}
\caption{The performance of the MAP vs. learned decoder with system parameter $(V = 3, M = 2048, D = 128)$ and $(V = 6, M = 2048, D = 256)$.}
\label{fig:s_map}
\end{figure}

Fig. \ref{fig:s_map} shows the BER performance comparison between the MAP decoder using eq. \eqref{eq:final_prob} and the DNN-based decoding. It indicates that the DNN decoder learns a decoding function that matches the performance of the approximated MAP decoding. Only slight degradation at high SNR is observed from the learned decoding scheme. For the remaining sections, we use the learned NOS encoder and the MAP decoder (eq. \eqref{eq:final_prob}) instead of a a learned NOS decoder because using (eq. \eqref{eq:final_prob}) has lower complexity and marginally lower BER. 

%% file: kbest.tex
\vspace{2mm}
\section{$K$-best Assisted Decoding}

In this section, we enhance the performance of learned NOS coding by applying a CRC-assisted $K$-best decoding algorithm to the MAP decoding output.

\subsection{Problem Formulation}
The neural network introduced in Section II is trained to 
minimize the number of errors per vector. However, typical mMTC applications do not tolerate any bit errors in a packet, hence the primary objective of our scheme is to minimize PER. For that, we include CRC bits in the information message to enhance the reliability of short packets in the low SNR regime. In our scenario, each NOS coded vector $\textbf{s}$ corresponds to a packet.

To introduce a K-best scheme, consider the joint probability $P(\mathbf{x}_1^{m_1},\cdots, \mathbf{x}_V^{m_V}|\y)$. Over all possible combinations of one-hot vectors $\{\mathbf{x}_1^{m_1}, \cdots, \mathbf{x}_V^{m_V}\}$, we desire to find the top-$K$ ($K$-best) combinations that maximize the joint probability. To efficiently evaluate the joint probability, we use eq. \eqref{eq:final_prob} in the form of the product of marginal probabilities given $\y$ based on the conditional independence assumption. 
\begin{align}
P(\mathbf{x}_1^{m_1},\cdots, \mathbf{x}_V^{m_V}|\y) = \prod \limits_{i=1}^V P(\mathbf{x}_i^{m_i}|\y).
\label{eq:joint prob}
\end{align} 

It is practically infeasible for large $V$ and $M$ to identify the exact $K$-best candidates via brute-force or tree-based searching. Thus we propose a $K$-best algorithm for simplified searching using the log-probability representation to replace the marginal probability multiplications with additions:
\begin{equation}
    \begin{split}
    log(P(\x_1^{m_1}, \cdots, \x_V^{m_V}|\y)) = \sum_{i=1}^V \mathbf{l}_i[m_i],
\label{eq:log prob}
    \end{split}
\end{equation}
where $\mathbf{l}_i[m_i]$ is the log-probability for $\x_i[m_i] = 1$.

\subsection{Proposed CRC-assisted $K$-best Decoding}

\begin{figure}[t]
\centering
\includegraphics[width=\linewidth]{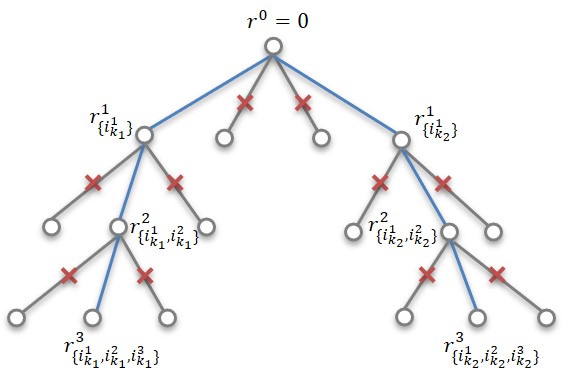}
\vspace{-0.3cm}
\caption{The proposed $K$-best algorithm. The two blue branches denote the $k_1$-th and $k_2$-th survived path for the first three layers.}
\label{fig:kbest}
\end{figure}

We adopt the general principle of $K$-best tree searching and pruning algorithms investigated in \cite{kbest} and \cite{kbest2} for near-maximum-likelihood MIMO detection. The objective is to find $K$-best candidates using a tree (Fig. \ref{fig:kbest}) to maximize the log joint probability \eqref{eq:log prob}. For the search, we define a score metric $r^v$ initialized as $r^0 = 0$ in a recursive form of: \\
\vspace{-3mm}
\begin{align}
r^v_{m_1,\cdots,m_v} = r^{v-1}_{m_1,\cdots,m_{v-1}}+\mathbf{l}_v[m_v].
\label{eq:update rule}
\end{align} 
Note $v$ indicates the index of $\textbf{x}_v$ or the level of the tree at which the score metric is evaluated. Unlike conventional $K$-best algorithms in MIMO detection, the update term of the score metric at the $v$-th layer, $\mathbf{l}_v[m_v]$, does not depend on the previous terms $\mathbf{l}_i[m_i], i \in [1, v-1]$ because of the conditional independence assumption. 
This property greatly reduces the complexity of $K$-best search compared to prior algorithms.

At each layer in the tree structure shown in Fig. \ref{fig:kbest}, we only keep $K$-best candidates by pruning out all the other nodes. For a $k$-th ($k\in [1,K]$) survived node in the $(v-1)$-th layer, we evaluate all metrics for its children nodes with index $m_v^k \in [1,M]$ based on eq. \eqref{eq:recursive}:
\begin{align}
r^v_{m^k_1,\cdots,m^k_v} = r^{v-1}_{m^k_1,\cdots,m^k_{v-1}} + \mathbf{l}_{v}[m^k_v] = \sum_{c=1}^{v-1} \mathbf{l}_{c}[m^k_c] + \mathbf{l}_{v}[m^k_v].
\label{eq:recursive}
\end{align}
In this way, $K\times M$  metrics are obtained at each layer and we only preserve the top-$K$ candidates to serve as the parent nodes for the next layer whereas all other nodes are pruned from the tree.  By repeatedly extending and pruning the $K$-best tree, $K$ survived paths are obtained at the last layer. The accumulated indices from the layer 1 to $V$ of the $k$-th survived path are denoted by  $m^k_1,\cdots,m^k_V$. By converting each $m^k_v$ to a bit sequence $\mathbf{b}^k_v$ and concatenating them together, we obtain the bit sequence $\mathbf{b}^k$ as the $k$-{th} candidate of the decoded bits.


The candidate bit sequence with the highest score metric serves as the MAP solution. However, with low SNR, the MAP solution may deviate from the correct bit sequence $\mathbf{b}$. To allow error detection and correction, the proposed $K$-best scheme produces a list of $K$-best candidates increasing the probability to have the correct $\mathbf{b}$ among these candidates. The last step is to check all candidates sequentially and select the first one which passes the CRC as the final result. The entire CRC-assisted $K$-best decoding algorithm for the learned NOS code is summarized in Algorithm \ref{algorithm}.
\begin{algorithm}
    
	\caption{$K$-best decoding CRC for NOS code.}
    \label{algorithm}

	\SetKwInOut{Input}{Input}\SetKwInOut{Output}{Output}
	\SetKwFunction{KBest}{KBest}
	\SetKwFunction{CRCcorrection}{CRCcorrection}
	\SetKwFunction{DecideOrder}{DecideOrder}
	\SetKwFunction{IdxToBits}{IdxToBits}
	\SetKwFunction{CRCDecode}{CRCDecode}
	\SetKwFunction{Reorder}{Reorder}
	\SetKwFunction{ChooseLayer}{ChooseLayer}
	\SetKwFunction{SelectNodes}{SelectNodes}
	\SetKwData{mylist}{outputList}
	\SetKwData{idx}{idx}
	\SetKwData{errFlag}{errFlag}
	\SetKwData{decodedBits}{decodedBits}
	\SetKwData{anc}{anc}
	
	\Input{$ K,\; \mathbf{l}_i \text{ for }i = 1, \cdots, V$}
	\Output{\decodedBits, \errFlag} 
	\fontsize{9}{9}\selectfont
	\BlankLine
	\For{$k=1$ \KwTo $K$}{
		
		$r^0(k) \leftarrow 0$ \hfill (zero score metric)\\
		$\idx(k) \leftarrow [\;]$ \hfill (empty candidate index)\\
	}
	
	\For{$v=1$ \KwTo $V$}{
	    $\mathbf{u} = \mathbf{l}_v$\\
		\For{$k=1$ \KwTo $K$}{
			$\mathbf{r}^{tmp}(k) \leftarrow r^{v-1}(k) + \mathbf{u}$\\
		}
		$[\mathbf{r}^v, \idx_{new}, \anc] \leftarrow$ \SelectNodes{$\mathbf{r}^\text{tmp}, K$}\\
		\For{$k=1$ \KwTo $K$}{
			$\idx(k) \leftarrow [\idx (\anc(k)), \idx_{new}(k)]$\\
		}
	}
	
	\While{\errFlag $\neq 0$ and $k \le K$}{
		\decodedBits $\leftarrow$ \IdxToBits{\mylist$(k)$}\\
		\errFlag $\leftarrow$ \CRCDecode{\decodedBits}\\
	}
\end{algorithm}

%

%% file: evaluation.tex
\section{Evaluation}

The PER performance of the proposed scheme is evaluated via Monte-Carlo simulations. The results show that the learned NOS code outperforms both HDM \cite{HDMkbest} and B/QPSK modulation protected by a 3GPP CRC-aided Polar code with a list decoding method \cite{3GPP_polar,polarld} for short packet transmission.

\subsection{Training Deep Learning Model}
Various DNN models for the encoder and decoder pair with different parameter sets $(V,M,D)$ are trained for $8\times 10^3$ epochs where $5\times 10^5$ training samples are used for each epoch. The batch size is chosen to be 1024 and a dynamic learning rate is adopted changing linearly from the initial value of $2\times 10^{-4}$ to the final $2\times10^{-6}$. All models are trained under a SNR fixed at $-1.5$dB while evaluated under different SNRs. 

\subsection{PER Performance}
\begin{figure}[t]
\centering
\includegraphics[width=\columnwidth]{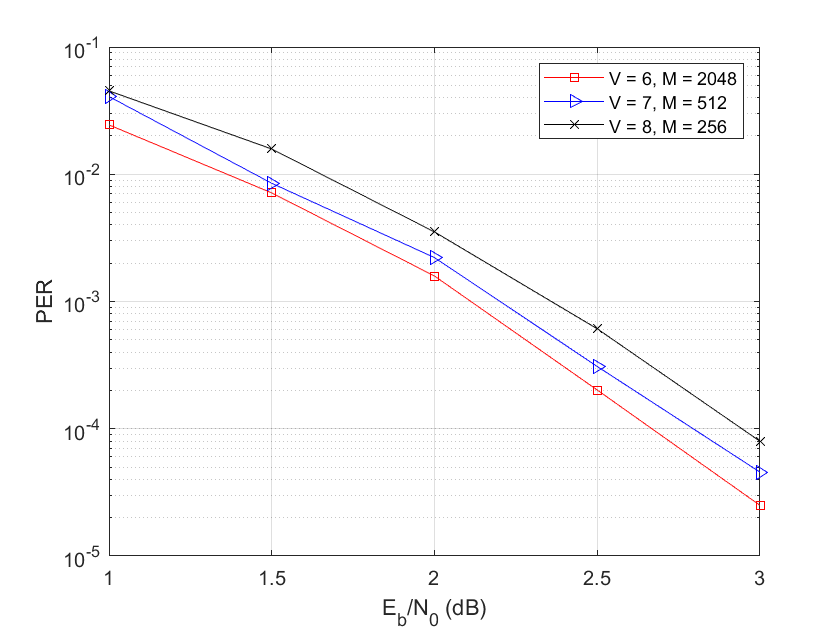}
\vspace{-0.3cm}
\caption{PER performance with different combinations of $(V,M)$. The curves are $(V = 8, M = 256)$, $(V = 7, M = 512)$ and $(V = 6, M = 2048)$ from the top to the bottom.}
\label{fig:VM}
\end{figure}

The rate $R$ of the proposed coding scheme is determined by the information length ($V \times m$-bit) and the length ($D/2$) of the complex-valued codeword, satisfying $R = \frac{V \times m}{D/2}$. Given $D$ and a target rate $R$, the selection of $(V,M=2^m)$ parameters has significant impact on the PER performance. To evaluate different parameter sets, consider a short packet with $D = 256$ and the target rate of $R \approx1/2$ (i.e., $\approx$64-bit message) which can be transmitted with different parameter combinations such as $(V = 8, M = 256)$, $(V = 7, M = 512)$, and $(V = 6, M = 2048)$. The PER performance of these parameter sets is shown in Fig. \ref{fig:VM}, where SNR is the ratio between the energy per information bit $E_b$ and the noise power $N_0$. Although the configuration of $(V = 6, M = 2048)$ supports the highest rate $R=0.5156$, it exhibits the lowest PER.

\begin{figure}[t]
\centering
\includegraphics[width=\columnwidth]{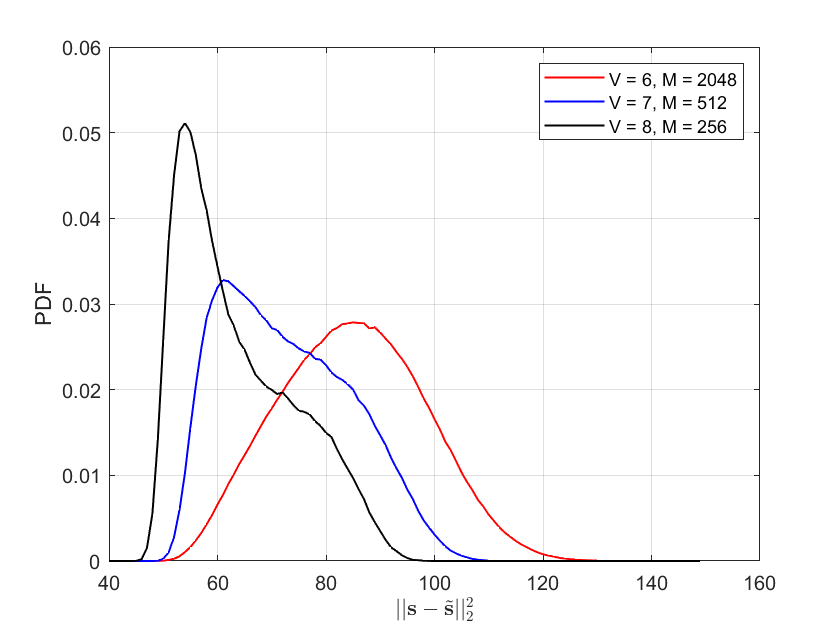}
\vspace{-0.3cm}
\caption{$||\textbf{s}-\tilde{\textbf{s}}||_2^2$ with different combinations of $(V,M)$, namely $(V = 6, M = 2048)$, $(V = 7, M = 512)$ and $(V = 8, M = 256)$.}
\label{fig:dist}
\end{figure}

Fig. \ref{fig:VM} shows the trend that the learned NOS coding performance enhances with a smaller $V$ and a larger $M$ for the same rate. This is also confirmed in Fig. \ref{fig:dist} where we plot the distribution of the pairwise distance $||\textbf{s}-\tilde{\textbf{s}}||_2^2$ among transmit vectors for different  $(V,M)$ settings. The neural network tends to learn near-orthogonal codewords for different encoders $Enc_i$ and it also learns to increase the distance among codewords from the same encoder. We observed that the minimum pairwise distance is mostly determined by the two nearest codewords generated by the same encoder (the near-orthogonal property makes the codeword distance from different encoders larger). When $V$ is relatively small, there are fewer encoders while each one is trained with a larger $M$ with more degrees of freedom (larger input with a bigger network) to learn a good coding scheme. Thus, we observe that a smaller $V$ with a larger $M$ for the same rate is generally preferred to achieve better PER as it allows the network to learn a better scheme with a larger minimum pairwise distance. 



Although a smaller $V$ with a larger $M$ improves the PER, it comes at the cost of increased network complexity. When $V$ is set to 4 for the scenario of $R=1/2$ and $D = 256$, $M$ needs to be as large as $2^{16}$. And the network size becomes quickly impractical to train when $V$ is less than 4. Note that $V=1$ is a special case to learn a single network to encode the entire bit-sequence without superposition, which is practically infeasible even for a relatively short packet length (e.g., 32-bit input requires a network with the input size of $M=2^{32}$). For practical mMTC applications, $V\geq3$ is a reasonable choice.

\begin{figure}[t]
\centering
\includegraphics[width=\columnwidth, height=180pt]{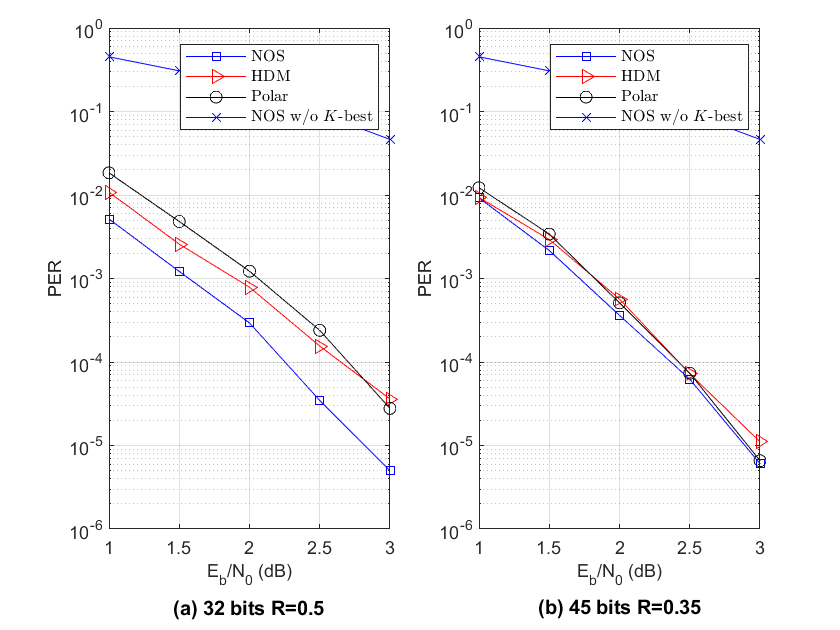}
\vspace{-0.3cm}
\caption{Comparison between the NOS code, HDM, and Polar code with CRC. (a): 33 info. bit transmission with NOS $(V=3, M=2048, D=128)$ vs. 32-bit with HDM and Polar. Spectral efficiency is $\approx0.5$ for all schemes. (b):  44-bit transmission with NOS $(V=4,M=2048,D=256)$ vs. 45-bit with HDM and Polar. Spectral efficiency is $\approx0.35$ for all schemes. All schemes use the same CRC.}
\label{fig:subplot}
\end{figure}

Finally, we compare the learned NOS code with HDM and Polar code to transmit short packets. In one case, we evaluate 32 or 33 message bits transmission (Fig. \ref{fig:subplot} (a)), and in the other case we increase the message length to 44 or 45 bits (Fig. \ref{fig:subplot} (b)). NOS coding uses the parameter set $(V = 3, M = 2048, D = 128)$ for the first case (33-bit) and $(V=4,M=2048,D=256)$ for the second case (44-bit). $K$-best decoding in both cases uses $K = 128$. HDM parameters are selected to transmit 32 ($R=0.5$) and 45 ($R=0.35$) message bits for respective cases with the $K$-best decoding scheme proposed in \cite{HDMkbest} using $K=128$. The 3GPP Polar code \cite{3GPP_polar} uses QPSK with coding rate of $0.25$ and $0.176$ for 32 and 45 message bits, respectively. Polar decoding uses CRC-assisted successive cancellation list decoding with the list size of $L=8$. Note that all these schemes have a similar spectral efficiency of $\approx0.5$ and $0.35$ bits/Hz/sec for the first and second case, respectively. The same 11-bit CRC is adopted to all schemes. Because of the near-orthogonality property, the proposed NOS decoding has significantly lower complexity (and shorter run-time) than HDM and Polar decoding on a desktop CPU. 


Fig. \ref{fig:subplot} (a) shows that the learned NOS scheme outperforms both HDM and Polar coding schemes with approximately 0.5dB gain for the same PER for 32-bit transmission. It also shows significant gain of the proposed CRC-assisted $K$-best decoding algorithm for our NOS scheme. The difference between our NOS scheme and HDM arises from the way to generate/learn superimposed codewords with less interference. HDM employs FFT with random permutations \cite{HDMkbest} without a structure to attain near-orthogonality among superimposed codewords. On the other hand, our scheme implicitly learns near-orthogonal codewords via joint training of the DNN-based encoder and decoder pair. 

When the number of message bits and the packet (codeword) length $D$ grow, the gain of NOS code gradually diminishes as shown in Fig. \ref{fig:subplot} (b). Diminishing performance gap is expected because Polar codes approach capacity-achieving performance as the length increases. On the other hand, longer message lengths make the size of the learned codebook ${\cal C}$ grow exponentially for the learned NOS code, and its parameter $V$ needs to increase (resulting in performance degradation) to maintain practical DNN training complexity. Nevertheless, the proposed NOS scheme is a promising solution for reliable short message transmission in the low SNR regime, especially for short packets. Investigating a more efficient DNN structure and its training scheme for longer information bit lengths is left as future work.